# Simultaneous pixel detection probabilities and spatial resolution estimation of pixelized detectors by means of correlation measurements

V. Grabski *

*Instituto de Física, Universidad Nacional Autonoma de Mexico, A.P. 20-364, 01000 DF, Mexico*

## Abstract

On the basis of the determination of statistical correlations between neighboring detector pixels, a novel method of estimating the simultaneous detection probability of pixels and the spatial resolution of pixelized detectors is proposed. The correlations are determined using noise variance measurement for isolated pixels and for the difference between neighboring pixels. The method is validated using images from two image-acquisition devices, a General Electric Senographe 2000D and a SD mammographic unit. The pixelized detector is irradiated with X-rays over its entire surface. It is shown that the simultaneous pixel detection probabilities can be estimated with an accuracy of 0.001–0.003, with an estimated systematic error of less than 0.005. The two-dimensional presampled point-spread function ($PSF^0$) is determined using a single Gaussian approximation and a sum of two Gaussian approximations. The results obtained for the presampled $PSF^0$ show that the single Gaussian approximation is not appropriate, and the sum of two Gaussian approximations providing the best fit predicts the existence of a large (~50%) narrow component. Support for this observation can be found in the recent simulation study of columnar indirect digital detectors by Badano et al. The sampled two-dimensional PSF is determined using Monte Carlo simulation for the L-shaped, uniformly distributed acceptance function for different fill-factor values. The calculation of the presampled modulation transfer function based on the estimated $PSF^0$ shows that the observed data can be reproduced only by the single Gaussian approximation, and that when the sum of two Gaussians is used, significantly larger values are apparent in the higher-frequency region for images from both detection devices. The proposed method does not require a precisely, constructed tool. It is insensitive to beam collimation and to system physical size and may be indispensable in cases where thin absorption slits or edges are difficult to use. It could therefore be very useful for regular detector verification. 

* Corresponding author. Tel.: 52-555-622-5185; fax: 52-555-622-5009; e-mail: Varlen.Grabski@cern.ch, grabski@fisica.unam.mx.

## 1. Introduction

One of the most common methods of determining the spatial resolution of pixelized detectors is measurement of the line-spread function using a narrow slit [1]. This technique requires precise slit fabrication and beam alignment. A similar technique is the edge-spread-function method, which is easier to use [2]. Both these methods need precisely constructed tools and can introduce systematic errors if operating techniques are imperfect [1–6]; moreover, different methods can be used for the same equipment to estimate systematic errors [3,5]. The other source of systematic errors that should be mentioned is aliasing, which cannot be estimated easily [4]. The two-dimensional presampled point-spread function ($PSF^0$) measurements, obtained using a round-hole collimator [6] or cylindrical absorber, have similar technical difficulties.

An alternative to these methods is to evaluate the system response in a periodic pattern [4,5]. Using this method, the aliasing problem can be solved, but precise technical execution is still required. The results of these alternative methods [5] for obtaining the modulation transfer function (MTF) agree within 5%–10% for frequencies of 4–5 cycles/mm. All these methods are complex, geometry-dependent, and require specially constructed tools. Moreover, none of them is convenient for frequent verification of digital detector performance.

This study proposes an approach which is entirely different from the above-mentioned methods, and of which a short version has already been published [7]. This method is based on the use of noise variance measurements to estimate the correlations between pixels. The author considers simultaneous detection of two pixels to be the main reason that these correlations exist. The determination of the simultaneous detection probability allows determination of the two-dimensional symmetric $PSF^0$. For detectors using columnar scintillators, the previous statement should be considered as an approximation [8].

For this method, it is important to know the pixel acceptance function (the area dependence of the pixel response), in which the estimation of $PSF^0$ includes the choice of the best parameterization and the estimation of the free parameters. The number ($\geq 3$) of free parameters that can be estimated depends on the relationship between the pixel size and the standard deviation of $PSF^0$.

To apply this method, it is necessary to consider 20–30 flat images (i.e., images with uniform irradiation of the entire detector surface) for different exposure values to extract electronic and quantum noise. Using single or paired images, the noise variance over all pixels is determined. In this case, the averaging procedure will introduce some systematic error because the noise content of the pixels is unequal. By using a few thousand images and performing noise determination over all of them, this method allows the determination of the simultaneous detection probability almost without systematic errors.

The main difficulty is the determination of the free parameters in the $PSF^0$ parameterization, in which only four simultaneous probabilities are different from zero. This will restrict the number of free parameters to three (this problem is also common to slit and edge methods, where the number of significant points is limited [9]). In the author's opinion, this is the main limitation of this method for detectors in which the standard deviation of the presampled $PSF^0$ is smaller than half the pixel size. The other requirement for determination of the $PSF^0$ is knowledge of the pixel acceptance function, a requirement which is common to all methods. In this study uniform response over the entire photodiode area is used in the absence of other information. The sizes of the photodiodes and the distances between them were obtained from a drawing from General Electric Company [10].

Another product of this method is a set of simultaneous detection probabilities, which can be directly used for image restoration (restoring a degraded image by simultaneous-counting pixels).

This method has a number of advantages: it is not sensitive to beam collimation and geometry, it does not require precisely constructed slits, edges, or patterns, and it can be used frequently as a detector quality test. To validate the proposed method, images produced by GE Senographe digital mammography units were used.

## 2. Method

The proposed method is intended to determine the simultaneous detection probability and the coordinate resolution of a pixelized detector and is based on the correlations between neighboring elements. The correlations are determined by noise variance measurement for isolated pixels and the difference between neighboring pixels. In general, the variance of the distribution of differences between two random variables $N_{ij}$ and $N_{mn}$ can be written as [11]:

$$V(\Delta N_{ijmn}) = V(N_{ij}) + V(N_{mn}) - 2\rho_{ijmn}\sqrt{V(N_{ij})V(N_{mn})}, \quad (1)$$

where $V(N_{ij})$, $V(N_{mn})$, and $V(\Delta N_{ijmn})$ are the variances of $N_{ij}$, $N_{mn}$, and $\Delta N_{ijmn} = N_{ij}-N_{mn}$ respectively. $\rho_{imnj}$ is the correlation coefficient, which is equal to zero if $N_{ij}$ and $N_{mn}$ are independent. This relation can be easily applied to a single flat image from a pixelized detector if the image is ergodic and stationary. Ergodicity means that the probability density function (pdf) defined relative to a point and calculated at each point across an infinite image and the pdf calculated at a given point across several different images are equivalent. If ergodicity holds, then stationarity means that the pdf of an image does not depend on the image point [12]. In this case, $\Delta N_{ijmn}$ and $\rho_{ijmn}$ will depend only on the difference of $k = m-i$ and $l = n-j$. In general, for flat images, if pixels contain similar amounts of electronic noise, these two conditions can be satisfied. Thus, there is no need to consider thousands of images to determine the pixel variance. The correlation coefficients in Eq. (1) can then be determined by measuring the ratio $R_{kl} = V(\Delta N_{kl})/V(N)$ using a single image. For variance determination, it is recommended to use a single image if that image is ideally flat (Method I) and two images (Method II) if the images are not ideally flat. For a single image, $V(\Delta N_{kl})$ and $V(N)$ are determined as the square of the standard deviations (the Gaussian distribution is not always appropriate) of the distributions of $\Delta N^i_{kl}$ and $N^i$ respectively, constructed in a given area of a pixel detector. In the case of two images, $V(\Delta N_{kl})$ and $V(N)$ were determined as the square of the standard deviations of the distributions of $\Delta N^{i1}_{kl}$ - $\Delta N^{i2}_{kl}$ and $N^{i1}$ - $N^{i2}$ (where $\Delta N^{i1}_{kl}$, $\Delta N^{i2}_{kl}$ are the differences and $N^{i1}$, $N^{i2}$ are the pixel values for the first and the second images respectively). Using Method II, one should expect to eliminate the contribution of spatial variation to the variance as determined over all pixels [13]. In the general case, when the images are not ideally flat, the following relation for $R_{kl}$ can be written (see Appendix A):

$$R_{kl} = 2(1 - \rho^Q_{kl}(1 - \frac{V^E}{V} - \frac{V^F}{V}) - \rho^E_{kl}\frac{V^E}{V} - \rho^F_{kl}\frac{V^F}{V}), \quad (2)$$

where $\rho^Q_{kl}$, $\rho^E_{kl}$, and $\rho^F_{kl}$ are the correlation coefficients (see Appendix A), and $V$, $V^E$, and $V^F$ are total, exposure-independent (also called electronic), and spatial-variation (flatness) variances respectively. To determine $\rho^Q_{kl}$ using Eq. (2), it is necessary to determine all other terms on the right-hand side. The contribution of the last two terms can easily be determined by measuring $R_{kl}$ in regions where $\rho^Q_{kl}$ is zero. In general, $V^F$ varies as the square of the pixel response $N$, and this dependence can be important for large values of $N$. The determination of $R_{kl}$ using two different images will simplify Eq. 2, leaving only the term that includes $V^E$ (see Appendix A).

The determination of the $V^E$ component of the Senographe 2000D mammographic pixel-detector device was previously reported in [13]. In that study, $V^E$ was considered to be the part of the total variance that was independent of X-ray radiation and was extracted by fitting a second-order polynomial to the total variance data, $V^E$ being the constant term of the polynomial. The method suggested in [13] will be used for the extraction of $V^E$ and $V^F$ as the constant and quadratic terms respectively of Eq. 3:

$$\mathrm{V}(N) = a_0 + a_1 N + a_2 N^2. \quad (3)$$

The quantum noise variance is proportional to the number of photons; thus, $a_1$ is the scaling factor for the photon signal transition. The $a_2N^2$ term represents the spatial variation, with scaling factor $a_2$. All parameters should be determined by fitting the total variance data.

To determine $\rho^Q_{kl}$, it is necessary to estimate $R_{kl}$. The number of possible ways to construct the difference $\Delta N_{kl}$ between the nearest-neighbor and second-nearest neighbor pixels used for $R_{kl}$ determination is 24 ($k,l$ = 1, 2, where nonzero values of $\rho^Q_{kl}$ are expected). In this method, there is a symmetry $R_{kl} = R_{-k-l}$ which limits the number of possible ways to determine all the $\rho^Q_{kl}$. Thus, the same symmetry exists between $\rho^Q_{kl}$ values, and the number of correlation coefficients that can be determined is 12. The arrangement of these $\rho^Q_{kl}$ relative to the matrix structure of the pixelized detector [10] is shown in Fig. 1. A coordinate system connected with a given pixel center ($i,j$) is used. Here the nonzero value of $\rho^Q_{kl}$ can be explained by the simultaneous detection of two pixels.

The probability of simultaneous detection of two neighboring pixels ($i,j$) and ($m,n$) in the case of independence of the pixel coordinates will depend on the index difference, defined here as $\alpha_{kl}$. The relationship between $\alpha_{kl}$ and $\rho^Q_{kl}$ is described in Appendix B under the assumption that $\alpha_{kl}$ is negligible if the index difference is greater than two. This assumption is based on empirical correlation-coefficient values.

The reconstruction of the sampled PSF(r) using $\alpha_{kl}$ is based on solving Eq. C1 in Appendix C, which in general is a relatively difficult task. In the literature, PSF(r) is usually presented as a convolution of the presampled $PSF^0$(r) and the pixel acceptance function *P(r)*. To estimate $PSF^0$(r), the Gaussian distribution and the sum of two Gaussians are used [14]. The free parameters are determined using Eq. C1 (see Appendix C). For purposes of comparison with the experimental data, the corresponding MTF should be calculated, that is, the Fourier transform of $PSF^0$. The experimental data are related mostly to the one-dimensional MTF that can be easily determined using the two-dimensional function $PSF^0$(r) [14].

## 3. Results

### *3.1. Correlation coefficients*

The determination of correlation coefficients $\rho^Q_{kl}$ according to Eq. 2 assumes the estimation of $R_{kl}$, $V$, $V^E$, $V^F$, $\rho^E_{kl}$, and $\rho^F_{kl}$. For that purpose, flat phantom images were obtained from the flat-field 100-μm pixel detector (amorphous silicon below columnar CsI(Tl) scintillators) of GE digital-mammography units, property of the National Institute of Cancer Research, Mexico City. The raw images were taken using 28 kVp and 26 kVp Mo/Mo beams for two mammography devices, a GE Senographe, 2000D (*UN*_1) and an SD (*UN*_2) respectively. The images were acquired under two different sets of conditions: changing mAs values while keeping absorber thickness constant (UN_1 with absorber mAs (4–400), UN_2 without absorber mAs (5–40)), and varying the absorber thickness while keeping the mAs value constant (UN_1, 50 mAs). The beam collimator was set to produce a maximum field size (19.13x22.93 cm). A time interval of about 1 min between image acquisitions was used to reduce the effects of CsI phosphor memory [13].

The variances $V(\Delta N_{kl})$ and $V(N)$ in Eq. 1 were determined by Method I and Method II in the given area of the pixel detector. For images that are not ideally flat, both variances can depend on the area size and its location on the pixel detector. Fig. 2 shows $R_{kl}$ dependence on area size for the two detection units (for pixel mean value, 1900 on the raw data scale) using both methods. As can be seen in Fig. 2, for both detection devices and when Method II is used, $R_{kl}$ values are stable within 1% for areas larger than 1.5 × 1.5 cm (if the area is smaller, these statistical errors are larger). For both devices, when using Method I, $R_{kl}$ values decrease with increasing area size because of the contribution of spatial variation. This means that Method I can be applied only for small area sizes, and the optimum size for use of this method remains to be determined. The condition $R_{kl}$ = 2 for Method II mentioned in Appendix A is satisfied with accuracy better than 1% (statistical errors are approximately 0.5%) for areas larger than 3 × 3 cm.

To study the behavior of pixel variance at different locations in the pixel matrix, the author determined this variance for a number of 4 × 4 cm areas with different central coordinates. For each direction (10 or 01) in the pixel matrix, the area was scanned for three fixed central positions (Pos_1 = 300, Pos_2 = 1000, and Pos_3 = 1700). The detector size in the 10 and 01 directions is 1913 and 2293 pixels respectively. The dependence of pixel variances on the pixel index for three fixed scan positions is presented in Figs. 3(a) and 3(b). It can be seen that the flatness is different in the 10 and 01 directions. In the 10 direction, the variation across the whole surface is approximately 40%, and using Method I, one should expect approximately 4% systematic error for a 2 ×2 cm area. Later, for this reason, only Method II will be used for variance estimation.

The stability of $R_{kl}$ across the entire detector surface should depend on the flatness of the image. For the same positions, the calculated values of $R_{kl}$ are presented in Figs. 4(a) and 4(b). The values of $R_{05}$ and $R_{50}$ are fairly stable across the whole detector area, with variation about the mean of less than 0.5%. The deviation of $R_{05}$ and $R_{50}$ from 2 is less than 1%, which means that the large spatial variation of pixel variance mentioned earlier does not significantly affect $R_{kl}$. The values of $R_{10}$ and $R_{01}$ are not as stable as in the previous case. Note the slight increase near the center of the detector. This within-scan variation is approximately 1% and can be easily explained by the coordinate dependence of the spatial characteristics of the detector [8]. This 1% variation in $R_{kl}$ can introduce a variation of approximately 3%–4% in the correlation coefficients.

For the same conditions and for the area located around the detector center, $R_{0l}$ was calculated for different $l$ values. The results for the dependence of $R_{0l}$ on $l$ are presented in Fig. 5. The values of $R_{0l}$ for $l \geq 3$ are almost constant (close to 2). The deviation of $R_{0l}$ from 2 is less than 1% (normally ~0.5%, which is of the magnitude of the statistical error). It is also apparent that the $R_{0l}$ value for $l$ = 2 differs from two by 1% for Unit 2 and 2% for Unit 1, which can be explained by the existence of small correlations between non-nearest-neighbor pixels.

To determine $\rho^{Q}_{kl}$, it is also necessary to estimate the variances $V^E$ and $V^F$ for a given value of $N$. For the extraction of $V^E$ and $V^F$, the total variance dependent on the mean pixel response value was estimated. By curve fitting, parameters $a_0$, $a_1$, and $a_2$ were then determined as described in the section on method. Fig. 6 shows the dependence of the experimental data of $V(N)$ on the pixel mean value, obtained by use of two methods on the two detection devices with fitted curves. It is clear that the behavior of $V(N)$ as determined by Method II depends linearly on $N$. This means that Method II almost eliminates the contribution of spatial variation to $V(N)$. The values of the parameters and their errors obtained by fitting data from the two detection devices using Method II for the 4x4 cm area located close to the center of the detector (pixel index position 900x900, where both devices have approximately the same variance values) are presented in Table 1. As expected, the values of $a_2$ are close to zero, and the differences between the other parameters for the two devices are less than 5%. The results obtained here for $a_0$ up to 20% are larger than those reported by Burges [13]. This difference can be partly explained by the different $a_1$ and mAs values used. Precise estimation of $a_0$ is complicated and requires multiple images obtained under the same conditions. The method used is approximate and depends on the interval of mAs. For the mAs intervals (5–40) and (4–400) and for the 4x4 cm area with position 900x900, the same $a_0$ value is obtained (see Table 1), probably indicating that small mAs values are important for estimation of $a_0$ using Eq. 3. When $a_0$ is estimated using 50 fixed-mAs images and the absorber thickness is varied, the result is a 25% larger value for the same area position. Use of a fixed integration time for the image acquisition procedure provides an accurate way of measuring electronic noise, but this introduces some inconvenience for automatic noise determination. The values of $a_0$ estimated using Eq. 3 depend on the location of the area on the detector surface. $a_0$ values depend only weakly (~5%) on the pixel index in the 01 direction and show a monotonically decreasing relationship with the pixel index in the 10 direction. The variation in $a_0$ is approximately 50% from one side of the detector to the other in the 10 direction, which can not be explained only by the variation of scaling factor $a_1$. Using the coordinate dependence of the $a_0$, the variation inside the 4x4 cm area can be estimated (this same area size is used to estimate correlation coefficients). In the worst case, for small index values in the 10 direction, the variation of $a_0$ over this area is less than 15%, which decreases to 5% with increasing pixel index. The dependence of $a_0$ on the pixel index can probably be explained by the variation of $a_1$ and by the structure of the detector's electrical circuit.

There are two ways to determine $\rho^Q_{kl}$ after extracting the electronic variance and spatial-variation noise: one (the *Fit* method) by fitting Eq. 2 to values of $R_{kl}$ for different pixel mean values (where $\rho^Q_{kl}$ is a free parameter and $\rho^E_{kl}$ and $\rho^F_{kl}$ have been previously estimated by fitting Eq. 2 to $R_{kl}$ data for $k,l > 2$), and the other (the *Diff* method) by determining the two last terms of Eq. 2 for a fixed pixel mean value using measured $R_{kl}$ values from outside the correlation region ($\rho^Q_{kl}=0$).

The first method of determining $\rho^Q_{kl}$ is shown in Fig. 7, which shows the dependence of $R_{01}$ on the mean pixel value for the two detection devices. As it can be seen in Fig. 7, the description of $R_{01}$ data using Eq. 2 is acceptable. For Method II, the terms containing $V^F$ in Eq. 2 are insignificant because of the relatively small value of $a_2$. Results obtained for $R_{kl}$ are almost independent of area size over a large range of mean pixel values. Fig. 7 also shows the determination of $R_{kl}$ using Method I (to illustrate the flatness effect), where the variances are determined for a 1 × 1 cm area. The decrease of $R_{kl}$ with increasing mean pixel value indicates the contribution of the last term of Eq. 2. The nonzero values of $\rho^F_{kl}$ can probably be explained by the existence of locally flat regions. The results obtained for $\rho^Q_{kl}$ for both devices for the 4 × 4 cm area with center location at pixel indices 900x900 and the *Fit* method are presented in Table 2 (labeled as "*Fit*"). The use of a small number of pixels for the 2 × 2 cm area gives practically the same results for $\rho^Q_{kl}$, but with larger errors. The errors shown in Table 2 are parameter errors resulting from the fitting procedure.

The results of the second method for $\rho^Q_{kl}$ determination, notated as "*Diff,*" are illustrated in Fig. 8, where $\rho_{k1}$ values have already been estimated depending on $k,l$ for the two detection devices and for $N\approx1900$. All $\rho^Q_{kl}$ are zero within the errors of estimation when $k,l > 2$. The results obtained for $\rho^Q_{kl}$ for the same area location are also presented in Table 2. The mean values of $\rho^{Q}_{kl}$ for a given detector are almost the same, taking into account the error values. In practical terms, both methods are almost similar. The only difference is that the *Fit* method suggests a model to explain the small deviation of $R_{kl}$ from 2 for the *k,l* region where $\rho^{Q}_{kl} = 0$. The values obtained for $\rho^{Q}_{10}$ for fixed-mAs images have a small minimum (about ~5%) at pixel index location 900, and $\rho^{Q}_{01}$ is almost independent of the pixel index in the 10 direction for fixed pixel index in the 01 direction at location 900. From the $R_{kl}$ coordinate dependence in the 01 direction (see Fig. 4), one can expect something similar for $\rho^{Q}_{01}$ as in the previous case for $\rho^{Q}_{10}$.

The values obtained for $\rho^{Q}_{kl}$ for the two detection devices are slightly different, which can be partly explained by the difference in beam energy. The difference between $\rho^{Q}_{01}$ and $\rho^{Q}_{10}$ for the same device is about $3\sigma_d$ using the *Diff* method of calculation and about 6–7$\sigma_f$ for the *Fit* method. The observed difference between $\rho^{Q}_{11}$ and $\rho^{Q}_{-11}$ is of lesser significance. Other correlation coefficients are very small, to the point of being indistinguishable from zero. An explanation of the source of these differences will be given in Section 3.3.

### 3.2. Uncertainties of $\rho^{Q}_{kl}$ estimation

The uncertainties generated by the statistical and fitting procedures used are shown in the various tables and figures in this paper. Another type of error, called systematic error, contains uncertainties generated by the use of approximations in the proposed method and includes the following sources of error: the averaging procedure used for variance determination by estimation over all pixels (in the case where the pixel variances are not the same), the determination of $V^E$ and $V^F$, and the estimation of spatial and electronic noise correlations. In this case, $V^F/V$ is vanishingly small ($< 10^{-7}$) due to the small value of $a_2$ (see Table 1). Therefore, the contributions of spatial variation and spatial correlation to systematic uncertainty can be neglected.

For the upper limit of variation of the electronic noise variance, one can use the estimate provided in the previous section. As shown in Appendix A, this variation will introduce a systematic error of less than 1% into the determination of $\rho^{Q}_{kl}$.

The contribution of electronic noise correlations to the value of $R_{kl}$ can be estimated for certain *k,l* values (*k* or $l > 2$) when the value of $\rho^{Q}_{kl}$ is zero. This contribution can be estimated by the difference $2-R_{kl}$. Thus, for the *Diff* method, the systematic errors in the estimation of $\rho^{Q}_{kl}$ are conditioned mainly by the systematic errors in determination of $V^E$ (see Eq. 2) and also by the approximation that all pixels have the same noise variance.

The estimation results for $\rho^{Q}_{01}$ and $\rho^{Q}_{10}$ using the *Diff* method for the fixed- and variable-mAs images from unit 1 as a function of mean pixel value are shown in Fig. 9. A slight dependence on the mean pixel value can be observed. If there is a systematic error in $V^E$ estimation, then, using Eq. 2, it is possible to estimate this error as follows:

$$\rho_{kl}^{QM} = \rho_{kl}^{QR}\left(1 - \frac{\Delta V^E}{V - V^E}\right), \quad (4)$$

where $\Delta V^E$ is the systematic error in the estimation of $V^E$ and $\rho^{QM}{}_{kl}$ and $\rho^{QR}{}_{kl}$ are the estimated and actual correlation coefficients respectively. Considering that $\rho^{QR}{}_{kl}$ and $\Delta V^E$ are free parameters, it is possible to estimate them by fitting $\rho^{QM}{}_{kl}$ data obtained under the different image acquisition conditions illustrated in Fig. 9. The results obtained for $\rho^{QR}{}_{kl}$ for the images with fixed and variable mAs agree within 0.005 for all directions. The results for the fixed mAs images agree within a 0.001 error with the results obtained using the *Fit* method. The agreement between $\rho^{QM}{}_{kl}$ data obtained using images with fixed and variable mAs becomes worse (0.01) for $N < 500$. Thus the use of $\rho^{QM}{}_{kl}$ to estimate $\rho^{QR}{}_{kl}$ for $N > 500$ can introduce systematic errors as great as 0.01. Taking into account the approximate relation $\alpha \approx 0.5\rho$ (see the next section), the systematic error in determination of $\alpha$ is less than 0.005.

The estimate of $\Delta V^E/V^E$ is not stable, but is approximately 12% in the 01 direction for fixed mAs. In the other directions shown in Fig. 9, $\Delta V^E/V^E$ variation is less than 6%. Most of the data obtained for $\rho^{QM}{}_{kl}$ require the reduction of $V^E$ values (negative $\Delta V^E$) obtained using Eq. 3 (see Fig. 9).

The estimation of the correlation coefficient $\rho^E{}_{kl}$ was performed by fitting experimental data for various $k,l$ values ($k$ or $l > 2$) using Eq. 2. The values obtained are small ($\rho^E{}_{kl} \leq 0.05$), and the errors are similar or larger. The large errors in this estimation can be explained by the small values of $V^E/V$. For smaller values of $N$ (< 500) where one can expect the electronic noise contribution to increase, more accurate estimation of $\rho^E{}_{kl}$ is possible. Larger values of $\rho^E{}_{kl}$ are not observed, and the values obtained are in the above-mentioned region.

### *3.3. Spatial resolution*

The positive values extracted for $\alpha_{kl}$ using Eq. B4 (Appendix B) with values of $\rho^Q{}_{kl}$ (Table 2) are shown in Table 3. These approximate solutions, in the case where only the nearest neighbors are different from zero, are almost the same as those obtained using Eq. B5. The errors shown in Table 3 were estimated using Eq. B6. As can be seen, $\alpha_{kl}$ differs significantly from zero only for the nearest neighbors $k,l<2$ [7]. The difference between the two detection devices is the same as for the previously mentioned correlation coefficients.

Using the values obtained for $\alpha_{kl}$, the free parameters of $PSF^0$ were then estimated as described in Appendix C. For the pixel aperture area, an $L$-shape was used as illustrated in Fig. 1. Assuming that $d_{10}$ is equal to 100 μm, the characteristic sizes of the active pixel elements obtained from Fig. 1 are presented in Table 4. The pixel fill-factor value for this data set is 0.65. In the literature, another value (0.75) for the pixel fill factor was found [16]. This parameter is very important for extraction of the presampled PSF$^0$; for this reason, another data set was calculated by proportionally changing all sizes to get a fill factor of 0.75 (see Table 4). A quadratic shape also was used for the pixel area, keeping the character sizes for the *L* -shape given in Table 4.

It was assumed that inside the photodiode area, *P(r)* is a uniformly distributed function, and outside it is zero. The simulation results show that when using a quadratic shape for the photodiode area, it is difficult to explain the difference between the values obtained for $\alpha_{01}$ and $\alpha_{10}$. However, for the rectangular photodiode area obtained from Fig. 1, the simulation results are good enough to explain that difference. Using the sizes from Table 4, an asymmetry value of 10% between $\alpha_{11}$ and $\alpha_{-11}$ was obtained, which is slightly smaller than the observed 15%–20%. The L-shape introduced by up-down and left-right asymmetry of a few percent is impossible to determine by this method.

The free parameters of PSF$^0$ were estimated by use of the $\chi^2$ minimization procedure described in Appendix C. Standard programs were not used for this procedure in the interests of saving time. A simple approximate procedure was used to calculate $\chi^2$ values for the grid of free parameter values. The fitting procedure for the sum of two functions must be carefully carried out. There are many local minima, and it is important to ascertain that the minimum obtained is global. Fitting was performed in two stages: first, all local minima were found on a grid with a large step size, and then for each local minimum, the same procedure was repeated using a finer grid.

The hypothesis of a single Gaussian distribution for PSF$^0$, which follows from the observed MTF data, is not always appropriate for the data obtained using the *Diff* method. The values of $\chi^2$ obtained for different devices are in the 5–16 range (the probability that the value of $\chi^2$ is greater than 7.8 is 5% for three degrees of freedom, because the number of $\alpha_{kl}$ with different values is five). For the data obtained using the *Fit* method and with small errors, $\chi^2$ values are always larger than the limit just mentioned. The values obtained for $\chi^2$ using the sum of two Gaussians for PSF$^0$ are in the 5–12 range for the data with small errors and in the 0.5–1.2 range for the data with large errors. In this latter case, the probability limit is 3.8 for one degree of freedom. Therefore, the fitting results for the *Diff* data must be considered acceptable. The use of a single exponential distribution or an exponential-Gaussian combination for PSF$^0$ does not enable a better description of the data.

The two-dimensional sampled PSF and its projections on the 01 and 10 directions using the previously mentioned aperture function with fill factor 0.65 are illustrated in Figs. 10(a) and 10(b). As can be seen from Fig. 10(b), PSF is not a symmetric function. The estimated resolutions in the 01 and 10 directions are 41 μm and 42 μm respectively for Unit 1. The results obtained for the free parameters of PSF$^0$ with their corresponding $\chi^2$ values for different models, fill factors, and detection devices are presented in Table 5. This table also provides the corresponding resolution values calculated using the standard deviations of the sampled PSF projections in the 01 and 10 directions as $\sigma_r^2 = \sigma_{01}^2 + \sigma_{10}^2$. It is evident that the resolution for a given presentation of PSF$^0$ is independent of the fill factor. If the fill factor is changed, PSF$^0$ changes also. The resolution depends on the PSF$^0$ model and differs by approximately 6%–8% between the two models. The resolution obtained by the sum of two Gaussians differs from that calculated using MTF data [16] (~2% for Unit_1 and ~6% for Unit_2), and the difference between the resolutions for the two detection devices is approximately 8%. As has already been mentioned, this difference can be partly explained by the difference in beam energy.

It should be noted that this sum includes two different Gaussians having very different ν−s and approximately the same weights. This structure is probably related to the columnar structure of the detector X-ray converter. In the simulation study [8], the existence of a narrow component in $PSF^0$ was also related to light transport in the columns of the CsI converter. The value obtained for the standard deviation is slightly larger for the column with diameter ~10 μm. This result can be explained by the existence of a small layer of unstructured converter material between the photodiode and the columnar structure [8].

Using data from the two-dimensional presampled $PSF^0$, a one-dimensional MTF can be calculated [15] to compare with the existing data (see Fig. 11). The symbols in Fig. 11 show the GE-detector MTF measurements by the manufacturer and independent authors [16,17]. All these data can be described by a single Gaussian with ν = 34 for the GE data [16] and ν = 37–39 for the data examined in [17]. As expected, the values obtained here for MTF using a single Gaussian for $PSF^0$ are higher than the GE data (see Fig. 11). This can be explained by the fact that the data from the current experiments do not include the pixel-response variations conditioned by a non-100% fill factor.

Between the best-fit results and the empirical data, there is a large disagreement in the high-frequency region. If the MTF is calculated using only the wide part of $PSF^0$ (weighting factor of 1), then the agreement with the other data is sufficiently good (Fig. 11).

## 4. Discussion

Unfortunately, the process of acquiring images using commercial devices is very much a black-box operation. The detectors used here have a variation in pixel variance of 30%–40% over the whole surface, but fortunately the parameters of interest here are stable enough and are similar for both detectors. However, the author believes that the results obtained could be improved by analyzing better-quality (flat) images.

The correlation coefficients have a slight minimum around the detector centre. The coordinate-related variation of the correlation coefficients is approximately 3%. The same magnitude of variation can be expected for the simultaneous detection probabilities.

The observed difference between the correlation coefficients in the two perpendicular directions is significant. This difference between $\rho^Q_{01}$ and $\rho^Q_{10}$ can probably be explained by the rectangular shape of the pixels. The sizes presented in Table 4 are sufficient to explain that difference. Estimation using Monte Carlo simulation (assuming a radially symmetric $PSF^0$) shows that up-down and left-right asymmetry in the simultaneous detection probabilities can be a few percent, but this is impossible to determine with the method used here. The observed 3-4σ difference in α values for the two detection devices can be partly explained by the difference in the beam energy used for image acquisition.

The estimation of $PSF^0$ depends on the nature of the acceptance function. For the detectors under study, there are not enough points to fit independently all the parameters given in Table 4. The observed MTF data [16] indicate that a single Gaussian function for $PSF^0$ should be sufficient. However, the single Gaussian function does not provide an acceptable fit to the current experimental data. The results obtained for $PSF^0$ from the best fit are slightly distinguishable from the observed data. Previously published simulation work [8] inspires confidence that the results obtained for $PSF^0$ are believable. This can be explained by the existence of a very narrow Gaussian in $PSF^0$ ($\nu$~7–11 $\mu m$) which is probably difficult to measure using the edge and slit methods when the fill factor is less than 100%.

The method used for the determination of $PSF^0$ is based on the use of integral equations, which are sensitive to the choice of function type. It can be stated with confidence that a single Gaussian cannot describe the current data adequately. If the sum of two Gaussians is appropriate to describe $PSF^0$ (one describing the light deflection inside the central column and another for the light transmitted to the other columns), then a narrow Gaussian is necessary. If the existence of the narrow Gaussian in $PSF^0$ is real, then it can be measured by this method using a small-pixel-size thin-film transistor (TFT) matrix.

## 5. Conclusions

This research has proposed a novel method for the estimation of spatial resolution for a pixelized detector. This method provides an accurate estimation of the simultaneous detection probabilities, which can be useful for image restoration.

The presampled and the sampled two-dimensional $PSF^0$ distributions are estimated for different models and fill-factor values. With a single Gaussian approximation, it is not always possible to obtain acceptable fitting results. The use of the sum of two Gaussians provides better fitting results, suggesting the existence of a narrow component in $PSF^0$. Taking into account the simulation results of Badano et al., this result can be considered probable.

The observed MTF data can be reproduced using a single Gaussian approximation for $PSF^0$. The best-fit results for $PSF^0$ suggest larger values for the MTF in the high-frequency region.

The proposed method is simple and does not require any special, precisely constructed tools. This method is insensitive to physical size of the beam source and the system, which may be an indispensable feature in cases where a thin absorption slit or edge is difficult to use.

The method presented in this study does not require any human intervention, can be carried out in automatic mode, and could be very useful for regular detector verification.

## Acknowledgments

The author is grateful to M.E. Brandan for helpful discussions, to Y. Villaseñor, MD, for kindly providing access to the mammography unit, to radiological technicians of the National Institute of Cancer Research for technical support, and to M. Grabska for preparation of the manuscript.

## Appendix A

Let $N_i$ be a pixel response (considering the one-dimensional case to simplify the relation, where $i$ is the pixel index) expressed as a sum of two parts: one exposure-independent (called an electronic response, $N_i^E$) and the second a linearly exposure-dependent term ($N_i^Q$):

$$N_i = N_i^E + N_i^Q \,. \quad \text{(A1)}$$

In this case, the pixel variance calculated over all images is:

$$V(N_i) = E[(N_i^E + N_i^Q - \mu_i^E - \mu_i^Q)^2] = E[(N_i^{QE} - \mu_i^{QE})^2] = V(N_i^E) + V(N_i^Q)\,, \quad \text{(A2)}$$

where $E[]$ signifies the mathematical expectation and $\mu_i^{QE} = E[N_i^E] + E[N_i^Q] = \mu_i^E + \mu_i^Q$.

For flat images, $\mu_i^{QE}$ and $V(N_i)$ are constant over all pixels. Thus, the variance determination using a single image (calculated over all pixels) is adequate.

The variance of the difference $(N_i-N_j)$ can be written as a mathematical expectation of the value $(N_i-N_j)^2$:

$$\begin{aligned} V(N_i - N_j) &= E[(N_i-\mu)^2] + E[(N_j-\mu)^2] + 2E[(N_i-\mu)(N_j-\mu)] = V(N_i^{QE}) + V(N_j^{QE}) \\ &- 2E[(N_i^{QE} - \mu_i^{QE})(N_j^{QE} - \mu_j^{QE})] \end{aligned} \quad \text{(A3)}$$

Taking into account the independence of $N^E$ and $N^Q$, the last term of Eq. A3 can be simplified as stated in [11]:

$$E[(N_i^{QE} - \mu_i^{QE})(N_j^{QE} - \mu_j^{QE})] = \rho_{i-j}^Q \sqrt{V(N_i^Q)V(N_j^Q)} + \rho_{i-j}^E \sqrt{V(N_i^E)V(N_j^E)}\,, \quad \text{(A4)}$$

where $\rho_{i-j}^Q$ is the correlation coefficient between $N^Q_i$ and $N^Q_j$ and $\rho_{i-j}^E$ is the correlation coefficient between $N^E_i$ and $N^E_j$. Using Eqs. A3 and A4 and the relation $R_{ij} = V(N_i - N_j)/V(N)$, the result is:

$$R_{i-j} = 2(1 - \rho_{i-j}^Q(1 - V(N^E)/V(N)) - \rho_{i-j}^E V(N^E)/V(N))\,. \quad \text{(A5)}$$

In the case of non-flat images, the variance determination over all pixels is generally complicated. To obtain an approximation, a variable $N_i^F = \mu - \mu_i^{QE}$, characterizing flatness, can be introduced, where $\mu$ is the mean value of $N_i$ over the different pixels and is equal to the mean value of $\mu_i^{QE}$. In this case, the pixel variance $V^P$ over all pixels can be estimated as:

$$V^P(N) = V^A(N) + V(N^F)\,, \qquad \text{(A6)}$$

where $V^A(N) = \sum V(N_i)/n$, $n$ is the number of pixels in the area where $V^P$ is calculated, and $V(N^F)$ is the variance of $N^F$. For non-flat images and using the above approximation, Eq. A5 can be modified as:

$$R_{i-j} = 2(1-\rho^Q_{i-j}(1 - V^P(N^E)/V^P(N) - V^P(N^F)/V^P(N)) - \rho^E_{i-j}V^P(N^E)/V^P(N) - \rho^F_{i-j}V^P(N^F)/V^P(N))$$

, (A7)

where $\rho^F{}_{i-j}$ is the correlation coefficient between $N_i^F$ and $N_j^F$. To suppress the contribution of $N^F$ to the variance [13], the differences between the values of the same pixels for two different images can be used. In this case, the pixel variance calculated over all pixels for two different images can be written:

$$V^P(N_i^1 - N_i^2) = E[(N_i^1 - N_i^2)^2] = E[(N_i^1 - \mu_i)^2] + E[(N_i^2 - \mu_i)^2] +$$
$$+ 2E[(N_i^1 - \mu_i)]E[(N_i^2 - \mu_i)] = 2V^A(N_i) \qquad .$$

(A8)

The variance of the difference $(N_i\text{-}N_j)$ for two images can be written:

$$V^P(N_i^1 - N_i^2 - N_j^1 + N_j^2) = E[(N_i^1 - N_i^2)^2] + E[(N_j^1 - N_j^2)^2] - 2E[(N_i^1 - N_i^2)]E[(N_j^1 - N_j^2)]$$
$$= 2V^A(N_i^1) + 2V^A(N_j^2) - 2E[(N_i^1 - \mu_i - N_i^2 + \mu_i)]E[(N_j^1 - \mu_j - N_j^2 + \mu_j)] = 4V^A(N) - \qquad .\text{(A9)}$$
$$-2E[(N_i^1 - \mu_i)(N_j^1 - \mu_j)] - 2E[(N_i^2 - \mu_i)(N_j^2 - \mu_j)] = 4V^A(N) - 4E[(N_i - \mu_i)(N_j - \mu_j)]$$

Thus in this case, in the absence of correlation, the value of the $R_{ij}$ parameter is also independent of the flatness of the images. The covariance in Eq. A9 can be represented as:

$$E[(N_i - \mu_i)(N_j - \mu_j)] = \rho_{i-j}\sum_{i,j}^{n}\sqrt{V(N_i)V(N_j)}\,/\,n\,, \qquad \text{(A10)}$$

where $n$ has the same meaning as in Eq. (A6). If the pixel variances are not the same, then the use of $V^A$ instead of $V(N_i)$ will introduce a systematic error. This error in the determination of $\rho$ was estimated by simulation using uniform and Gaussian distributions up to a relative variation of 20% in the variance. The error value is almost independent of the form of the distribution and is about 1% for a 20% variation in variance.

Using thousands of images to calculate the variance along the different images, the determination of $R_{ij}$ by Eq. A5 can be shown to be mathematically correct.

## Appendix B

If the initial photon number in pixel $(i,j)$ is $\xi_{ij}$ and the simultaneous detection probability for the same photon is $\alpha_{mn}$ (where $m = \pm 0, \pm 1,..$ and $n = \pm 0, \pm 1,..$ ), then the real value $N_{ij}$ detected in pixel $(i,j)$ can be written (accounting for image degradation) as:

$$N_{ij} = \sum_{m,n=-s}^{s} \alpha_{mn} \xi_{i+mj+n} , \quad \text{(B1)}$$

where $s$ is the maximum number of pixels around a given pixel $(i,j)$ when $\alpha_{mn} \neq 0$. The relationship between the variances is:

$$V(N_{ij}) = \sum_{m,n=-s}^{s} \alpha_{mn}^2 V(\xi_{i+mj+n}) \quad \text{(B2)}$$

Assuming that all $\xi_{ij}$ are independent, then the covariance between pixel pairs $(i,j)$ and $(i + k, j + l)$ can be written as:

$$\begin{aligned} &\mathrm{cov}(N_{ij}, N_{i+kj+l}) = E[(N_{ij} - \mu_{ij}^{N})(N_{i+kj+l} - \mu_{i+kj+l}^{N})] = \\ &E[(\sum_{m,n=-s}^{s} \alpha_{mn}(\xi_{i+mj+n} - \mu_{i+mj+n}^{\xi}))(\sum_{u,v=-s}^{s} \alpha_{u,v}(\xi_{i+k+uj+l+v} - \mu_{i+k+uj+l+v}^{\xi}))] = \\ &\sum_{m=-s+k,n=-s+l}^{s} \alpha_{mn}\alpha_{m-kn-l} E[(\xi_{i+mj+n} - \mu_{i+mj+n}^{\xi})(\xi_{i+mj+n} - \mu_{i+mj+n}^{\xi})] = \sum_{m=-s+k,n=-s+l}^{s} \alpha_{mn}\alpha_{m-kn-l} V(\xi_{i+mj+n}) \end{aligned} . \quad \text{(B3)}$$

To simplify Eq. B3, the relations $m = u + k$ and $n = v + l$ (where m, u, n, and $v = 0, \pm 1,.., \pm s$), which follow from the independence of $\xi_{ij}$, can be used. Combining Eqs. (B2), (B3), and the relation between the covariance and the correlation coefficient $\rho_{kl}$ [11], an expression for $\rho_{kl}$ can be written as:

$$\rho_{kl} = \sum_{m=-s+k,n=-s+l}^{s} \alpha_{mn}\alpha_{m-kn-l} / \sum_{m,n=-s}^{s} \alpha_{mn}^2 , \quad \text{(B4)}$$

where $s$=2, which follows from the experimental data for the correlation coefficients of the detectors under study. Then the number of $\alpha$-s pairs is 25-1, because $\alpha_{00}$ can be normalized to one. In the suggested method, the number of correlation coefficients that can be measured is 12 for $s = 2$; to solve this problem, it can be assumed that $\alpha_{kl} = \alpha_{-k-l}$ (this is an approximation if the active pixel element has an L-shape). Thus, to determine α-s, a system of 12 quadratic equations (B4) must be solved. This system of equations is difficult to solve analytically for $\alpha$. Therefore, the standard program "c05nbc" from the Numerical Algorithms Group (NAG) library for solving nonlinear systems of equations was used [18]. In the case that only four correlation coefficients are different from zero, Eq. (B4) can be simplified to:

$$\rho_{01} = 2(\alpha_{01}, + \alpha_{10}(\alpha_{11} + \alpha_{-11}))/(1+2k)$$
$$\rho_{10} = 2(\alpha_{10}, + \alpha_{01}(\alpha_{11} + \alpha_{-11}))/(1+2k)$$
$$\rho_{11} = 2(\alpha_{11}, + \alpha_{10}\alpha_{01})/(1+2k)$$
$$\rho_{-11} = 2(\alpha_{11}, + \alpha_{10}\alpha_{01})/(1+2k) \qquad \text{(B5)}$$

where $k = \alpha^2_{01} + \alpha^2_{10} + \alpha^2_{11} + \alpha^2_{-11}$. This system of equations is also difficult to solve for $\alpha_{kl}$. However, (B5) can be approximately solved by removing members of order $O(\alpha^2)$:

$$\alpha_{01} = (2B_{01} - B_{10}(B_{11} + B_{-11}))/(4 - (B_{11} + B_{-11})^2), \qquad \text{(B6)}$$

where $B_{ij} = (1+2k)\rho_{ij}$. Taking into account that $k$ has a value between 0.02 and 0.03, the system of equations (B5) can be solved exactly for $\alpha_{kl}$ for a fixed value of $k$ (for example $k$ = 0.02):

$$16\alpha_{01}^2 - 8B_{01}\alpha_{01}^4 - 16\alpha_{01}^3 + (2(B_{11} + B_{-11})B_{10} + 8B_{01})\alpha_{01}^2 + (4 - (B_{11} + B_{-11})^2 - 2B_{10}^2)\alpha_{01} + \\ + (B_{11} + B_{-11})B_{10} - 2B_{01} = 0 \qquad \text{(B7)}$$

The values obtained for $\alpha_{kl}$ are used to calculate the new approximate value of $k$ for the subsequent iteration. After 10 iterations, the results obtained here agreed with the results of the standard program to within 1%. Other $\alpha$-s values can be easily determined using the system of equations (B5).

## Appendix C

The relation between $\alpha_{kl}$ and PSF can be written as:

$$\alpha_{kl} = \int_{\Omega_{kl}} PSF(r)dr \,/ \int_{\Omega_{00}} PSF(r)dr, \qquad \text{C1}$$

where PSF(r) should be determined as a convolution of the presampled $PSF^0(r)$ and a pixel-aperture function $P(r)$. $\Omega_{kl}$ is the area of the $(k,l)$ pixel (see Fig. 1), and $\Omega_{00}$ is the area of the central pixel. It can be expected that $PSF^0$ will be a symmetric function approximating a columnar scintillation X-ray converter [8].

Equation (C1) can be used for the evaluation of free parameters. For integration, the Monte Carlo method has been used. The free parameters of $PSF^0$ are determined by minimizing the following $\chi^2$ function:

$$\chi^2 = \sum_{k,l} (\alpha_{kl} - \int_{\Omega_{kl}} PSF(x,y)dxdy / \int_{\Omega_{00}} PSF(x,y)dxdy)^2 / \sigma^2_{\alpha_{kl}}, \qquad \text{(C2)}$$

where $\sigma_{kl}$ are the errors of $\alpha_{kl}$.

In the present study, a single Gaussian (from the observed MTF data) and a sum of two Gaussians (to detect possible contributions of the columnar structure of the *CsI* scintillator) have been used for $PSF^0$. For the single Gaussian function, there is one parameter to determine, and for the sum of two Gaussians, $PSF^0(r)$ has the following form:

$$PSF^{\,0}(r) = \frac{1-C}{4\pi\nu_1^2} e^{-\frac{r^2}{4\nu_1^2}} + \frac{C}{4\pi\nu_2^2} e^{-\frac{r^2}{4\nu_2^2}}, \qquad \text{(C3)}$$

where $\nu_1$, $\nu_2$, and $C$ are free parameters. The relationship between $\nu$ and the standard deviation $\sigma$ is $\sigma = \nu\sqrt{2}$. Thus, in this case, there are three free parameters that need to be determined using minimization procedure (C2).

Table 1. Fitting results for parameters $a_0$, $a_1$, and $a_2$ for different detection devices.

| Unit | area | $a_0$ | $a_1$ | $a_2$ |
|---|---|---|---|---|
| 1 | $2x2cm^2$ | 12.33 ± 0.13 | 0.148 ± 0.0004 | 1.0e-12 ± 6.5e-08 |
| 1 | $4x4cm^2$ | 12.20 ± 0.06 | 0.145 ± 0.0002 | 1.0e-12 ± 9.2e-09 |
| 2 | $2x2cm^2$ | 12.64 ± 0.22 | 0.147 ± 0.0005 | 1.0e-12 ± 5.5e-08 |
| 2 | $4x4cm^2$ | 12.46 ± 0.11 | 0.149 ± 0.0003 | 1.0e-12 ± 1.1e-08 |

Table 2. Correlation coefficients for different detection devices.

| | UN_1 Fit | UN_1 Diff. | UN_2 Fit | UN_2 Diff. |
|---|---|---|---|---|
| $\rho_{01}$ | 0.205 ± 0.002 | 0.210 ± 0.005 | 0.177 ± 0.002 | 0.185 ± 0.005 |
| $\rho_{02}$ | 0.013 ± 0.002 | 0.016 ± 0.006 | 0.000 ± 0.002 | 0.008 ± 0.006 |
| $\rho_{10}$ | 0.190 ± 0.002 | 0.191 ± 0.005 | 0.163 ± 0.002 | 0.169 ± 0.005 |
| $\rho_{20}$ | 0.005 ± 0.002 | 0.005 ± 0.006 | 0.000 ± 0.002 | 0.008 ± 0.006 |
| $\rho_{11}$ | 0.060 ± 0.002 | 0.063 ± 0.006 | 0.047 ± 0.002 | 0.047 ± 0.006 |
| $\rho_{22}$ | 0.001 ± 0.002 | 0.001 ± 0.006 | 0.000 ± 0.002 | 0.000 ± 0.006 |
| $\rho_{-11}$ | 0.070 ± 0.002 | 0.070 ± 0.006 | 0.056 ± 0.002 | 0.051 ± 0.006 |
| $\rho_{-22}$ | 0.000 ± 0.002 | 0.000 ± 0.006 | 0.000 ± 0.002 | 0.000 ± 0.006 |
| $\rho_{12}$ | 0.002 ± 0.002 | 0.006 ± 0.006 | 0.000 ± 0.002 | 0.001 ± 0.006 |
| $\rho_{21}$ | 0.005 ± 0.002 | 0.005 ± 0.006 | 0.000 ± 0.002 | 0.009 ± 0.006 |
| $\rho_{-12}$ | 0.005 ± 0.002 | 0.005 ± 0.006 | 0.000 ± 0.002 | 0.001 ± 0.006 |
| $\rho_{-21}$ | 0.008 ± 0.002 | 0.005 ± 0.006 | 0.000 ± 0.002 | 0.006 ± 0.006 |

Table 3. Simultaneous pixel detection probabilities for different detection devices obtained by use of correlation coefficients from Table 2.

| | UN_1 Fit | UN_1 Diff. | UN_2 Fit | UN_2 Diff. |
|---|---|---|---|---|
| $\alpha_{01}$ | 0.102 ± 0.001 | 0.105 ± 0.003 | 0.088 ± 0.001 | 0.092 ± 0.003 |
| $\alpha_{02}$ | 0.001 ± 0.001 | 0.002 ± 0.003 | 0.000 ± 0.001 | 0.001 ± 0.003 |
| $\alpha_{10}$ | 0.094 ± 0.001 | 0.097 ± 0.003 | 0.080 ± 0.001 | 0.083 ± 0.003 |
| $\alpha_{20}$ | 0.000 ± 0.001 | 0.000 ± 0.003 | 0.001 ± 0.001 | 0.000 ± 0.003 |
| $\alpha_{11}$ | 0.022 ± 0.001 | 0.023 ± 0.003 | 0.016 ± 0.001 | 0.016 ± 0.003 |
| $\alpha_{22}$ | 0.000 ± 0.001 | 0.000 ± 0.003 | 0.000 ± 0.001 | 0.001 ± 0.003 |
| $\alpha_{-11}$ | 0.027 ± 0.001 | 0.027 ± 0.003 | 0.020 ± 0.001 | 0.018 ± 0.003 |
| $\alpha_{-22}$ | 0.000 ± 0.001 | 0.000 ± 0.003 | 0.000 ± 0.001 | 0.001 ± 0.003 |
| $\alpha_{12}$ | 0.000 ± 0.001 | 0.001 ± 0.003 | 0.000 ± 0.001 | 0.000 ± 0.003 |
| $\alpha_{21}$ | 0.000 ± 0.001 | 0.000 ± 0.003 | 0.000 ± 0.001 | 0.000 ± 0.003 |
| $\alpha_{-12}$ | 0.000 ± 0.001 | 0.000 ± 0.003 | 0.000 ± 0.001 | 0.000 ± 0.003 |
| $\alpha_{-21}$ | 0.001 ± 0.001 | 0.000 ± 0.003 | 0.000 ± 0.001 | 0.001 ± 0.003 |

Table 4. Pixel photodiode character sizes (see Fig 1).

| | $d_{01}$ | $d_{10}$ | $\Delta_{01}$ | $\Delta_{10}$ | $l_{01}$ | $l_{10}$ | Fill factor |
|---|---|---|---|---|---|---|---|
| **From drawing** | **94** | **100** | **9.4** | **8.8** | **21.5** | **10** | **65%** |
| **From [15]** | **94** | **100** | **7.2** | **6.6** | **21.5** | **10** | **75%** |

Table 5. Free parameters and corresponding $\chi^2$ values obtained for different models and fill-factor values. Resolution determined using sampled PSF.

| | Fill fac. | $\nu$(μm) | $\chi^2$ | $\sigma_r$(μm) | C | $\nu_1$(μm) | $\nu_2$ (μm) | $\chi^2$ | $\sigma_r$(μm) |
|---|---|---|---|---|---|---|---|---|---|
| **Unit 1** | **0.65** | **30.9** | **13** | **53.8** | **0.55** | **10.7** | **39.8** | **1.2** | **58.5** |
| | **0.75** | **30** | **16** | **53.9** | **0.60** | **9.0** | **39** | **1.0** | **58.4** |
| **Unit 2** | **0.65** | **28.1** | **5.3** | **50.9** | **0.52** | **10.0** | **34.0** | **0.6** | **53.6** |
| | **0.75** | **27.2** | **6.6** | **51.2** | **0.60** | **7.5** | **34.0** | **0.6** | **53.9** |

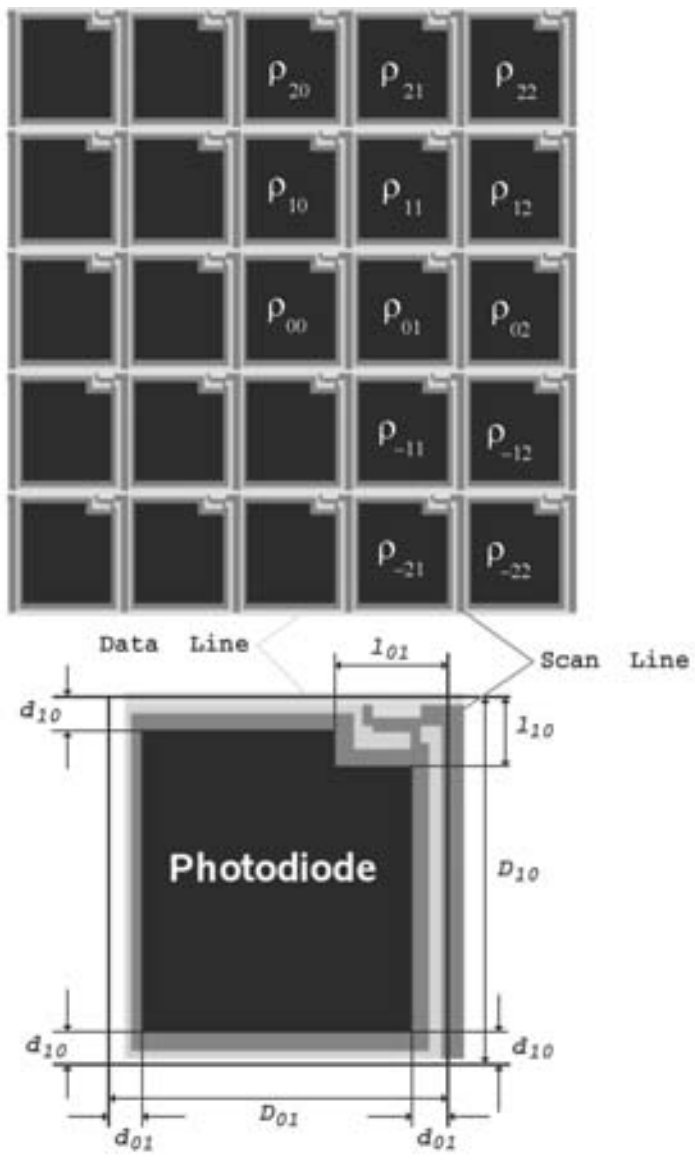

Figure 1. Pixel matrix structure. The pixel area sizes given in Table 4 are taken from [10]. The correlation coefficients defined in the text are shown.

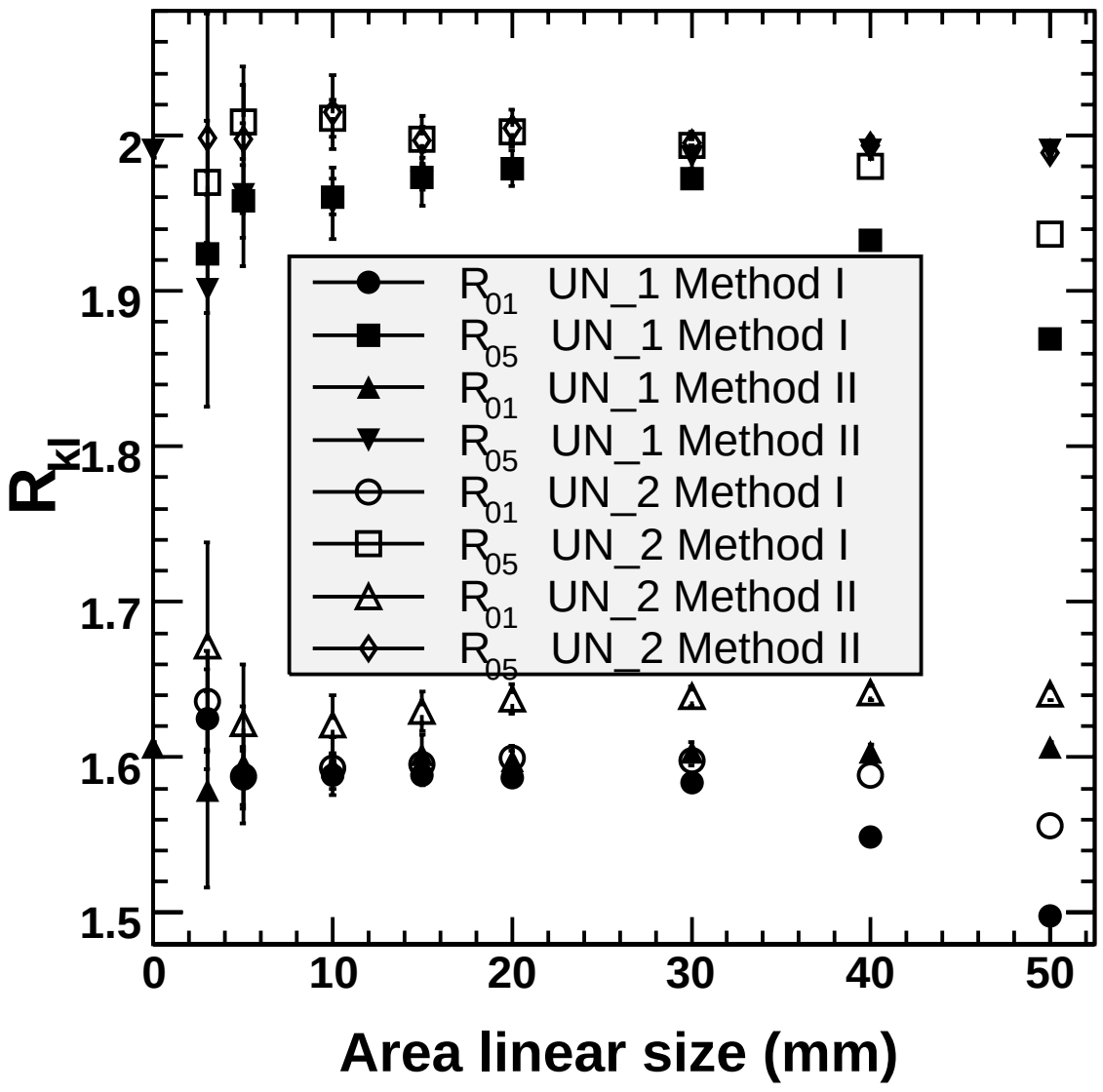

Figure 2. Area dependence of $R_{01}$ and $R_{05}$ for different detection devices and different methods of variance determination.

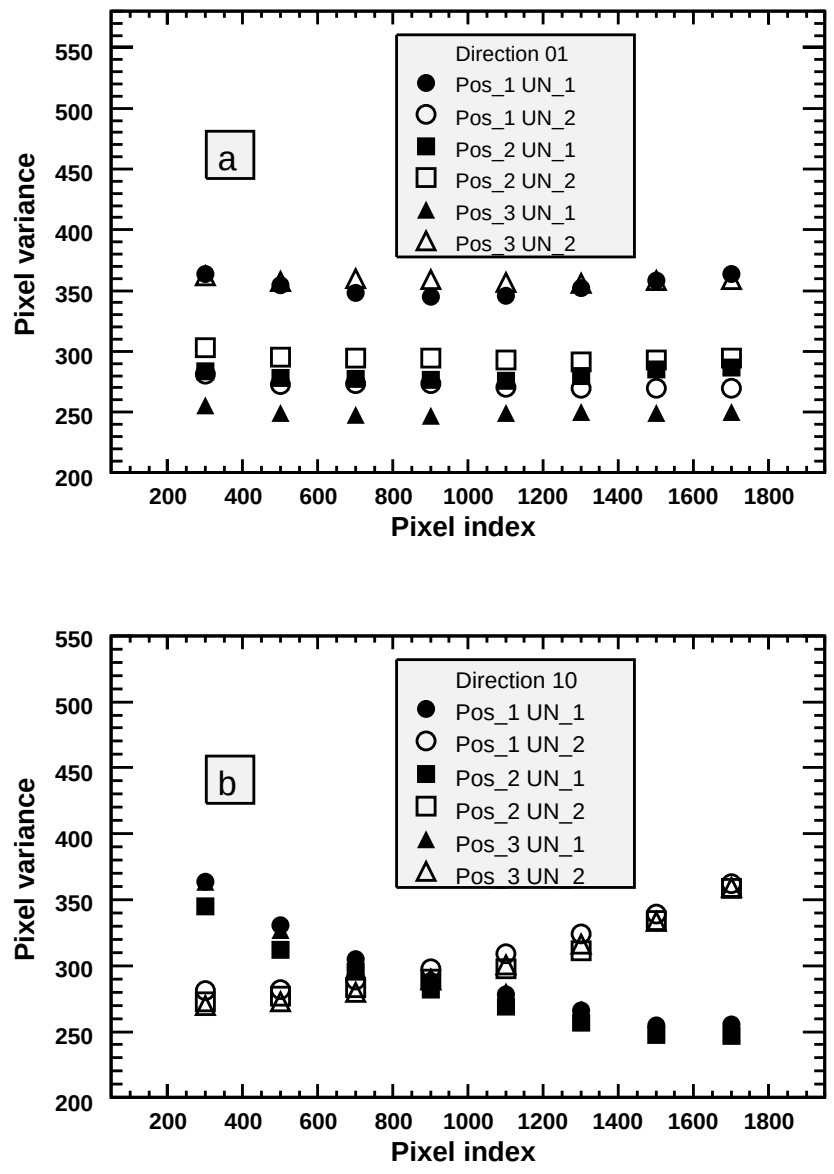

Figure 3. Pixel variance for the two detection units determined using method II on a $4 \times 4$ $cm^2$ area, for three different positions and two perpendicular directions dependent on the pixel index of the detector pixel matrix: (a) for the 01 direction and (b) for the 10 direction.

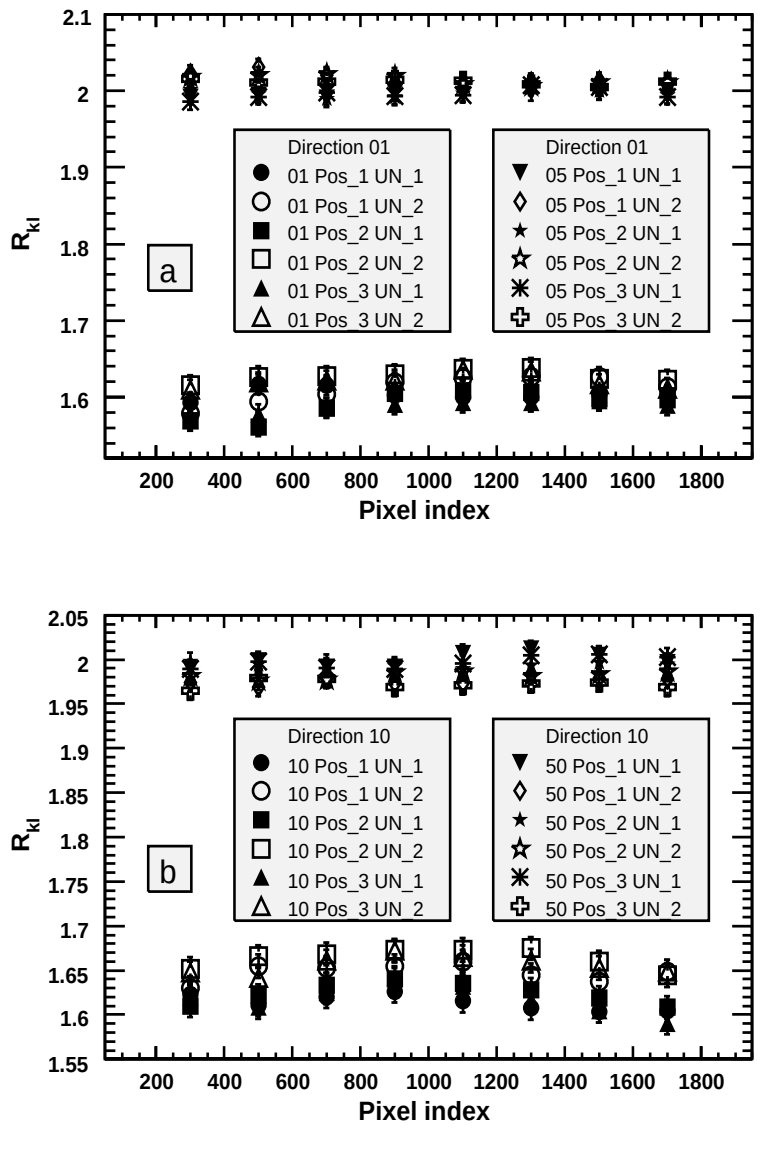

Figure 4. $R_{kl}$ dependence on the pixel index (same conditions as for Fig. 3).

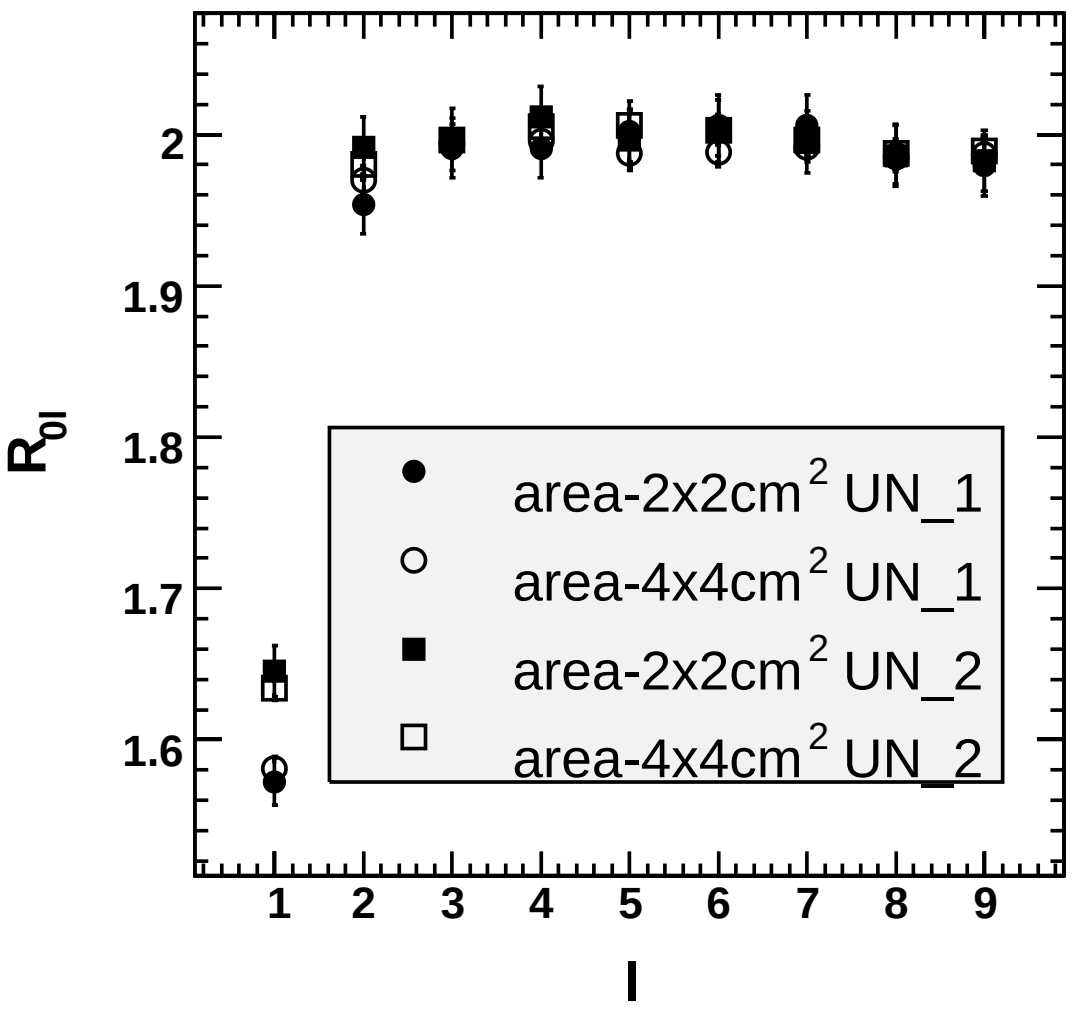

Figure 5. $R_{0l}$ dependence on the pixel index difference $l$ for different methods and detection devices.

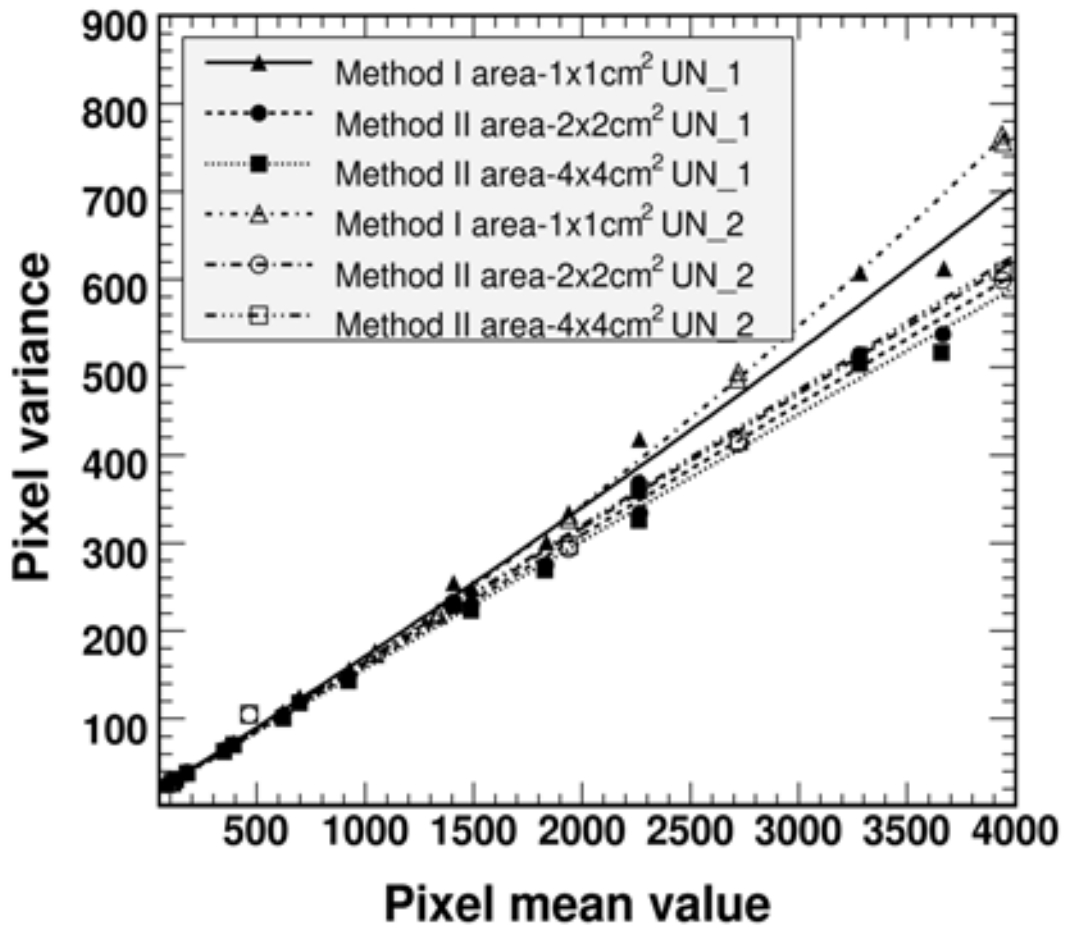

Figure 6. Pixel variance dependence on the pixel mean value for different methods and detection devices. Lines represent the results of fitting by the polynomial function in Eq. 3.

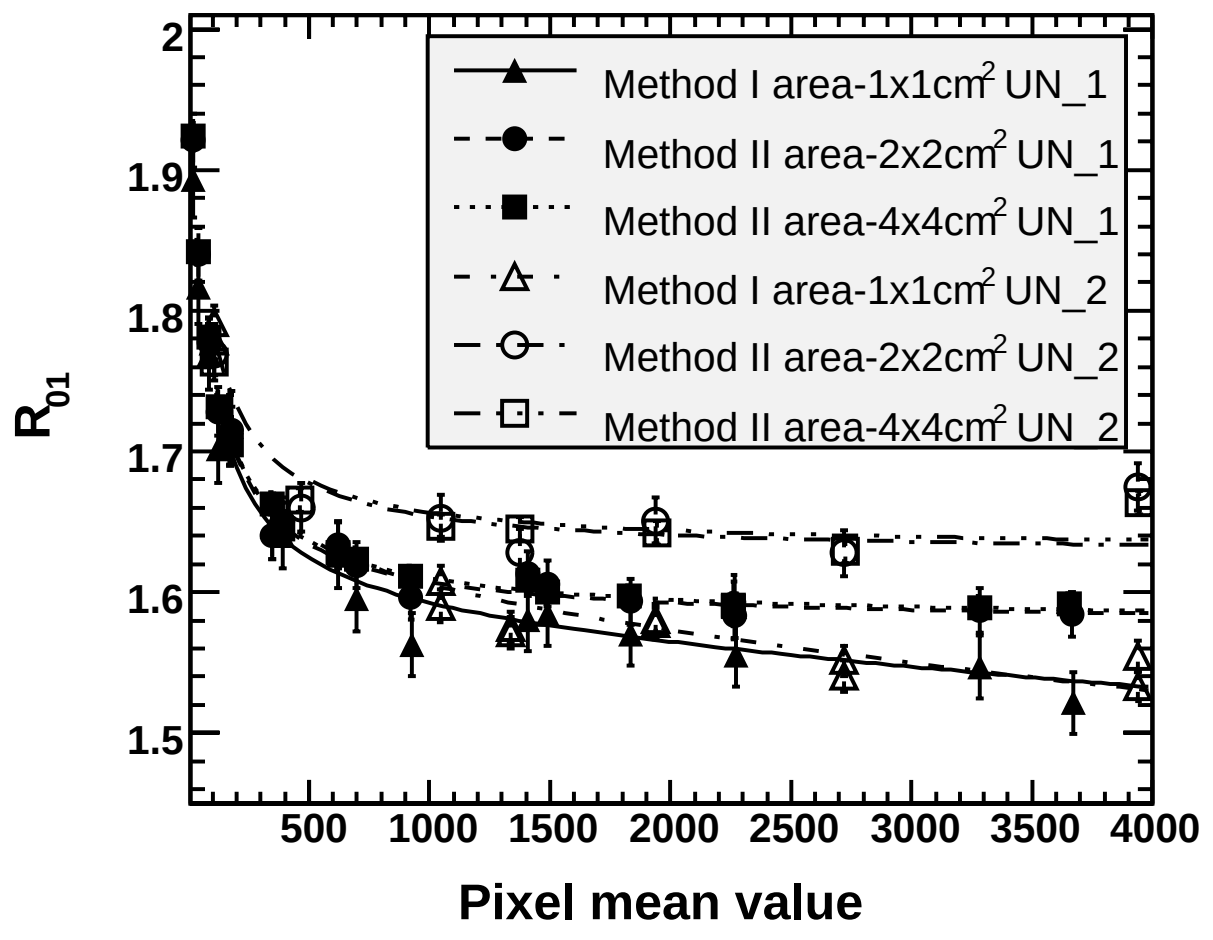

Figure 7. $R_{01}$ dependence on the pixel mean value for different detection devices and methods. Lines represent the results of fitting by Eq. 2.

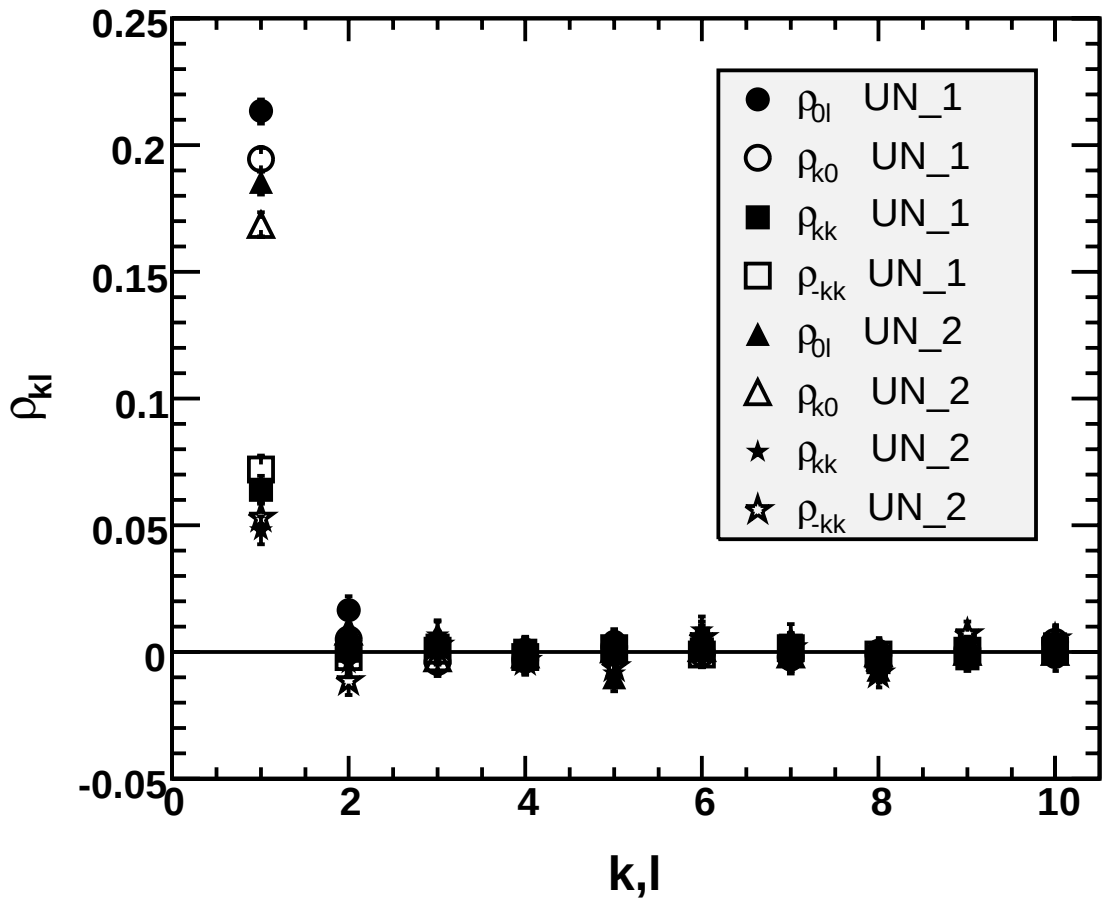

Figure 8. Dependence of correlation coefficients on the pixel index difference for different detection devices.

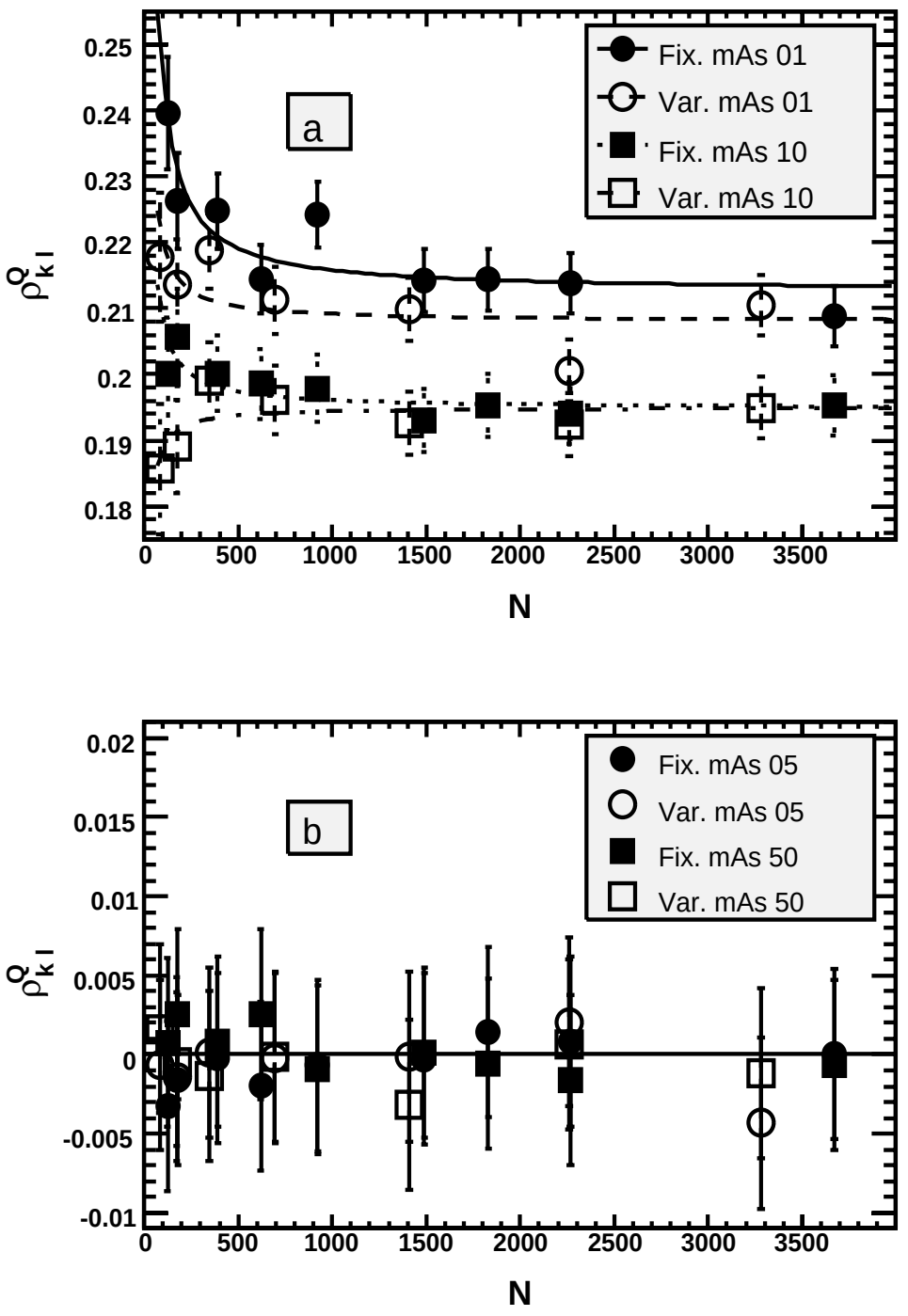

Figure 9 $\rho^{Q}_{kl}$ values estimated using Diff method for the fixed and varied mAs images. For the $\rho^{Q}_{01}$ and $\rho^{Q}_{10}$ (a) and for the $\rho^{Q}_{05}$ and $\rho^{Q}_{50}$ (b). Lines are fit results by Eq 4.

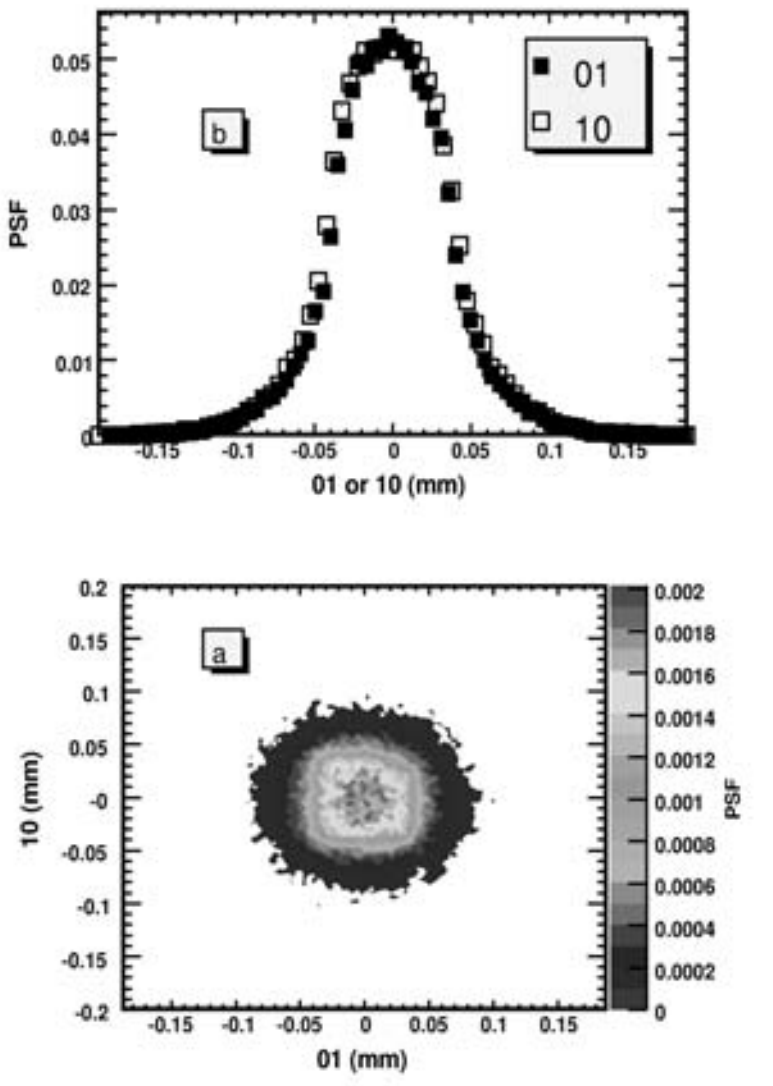

Figure 10. Two-dimensional sampled PSF (a) and its two projections in the 01 and 10 directions (b).

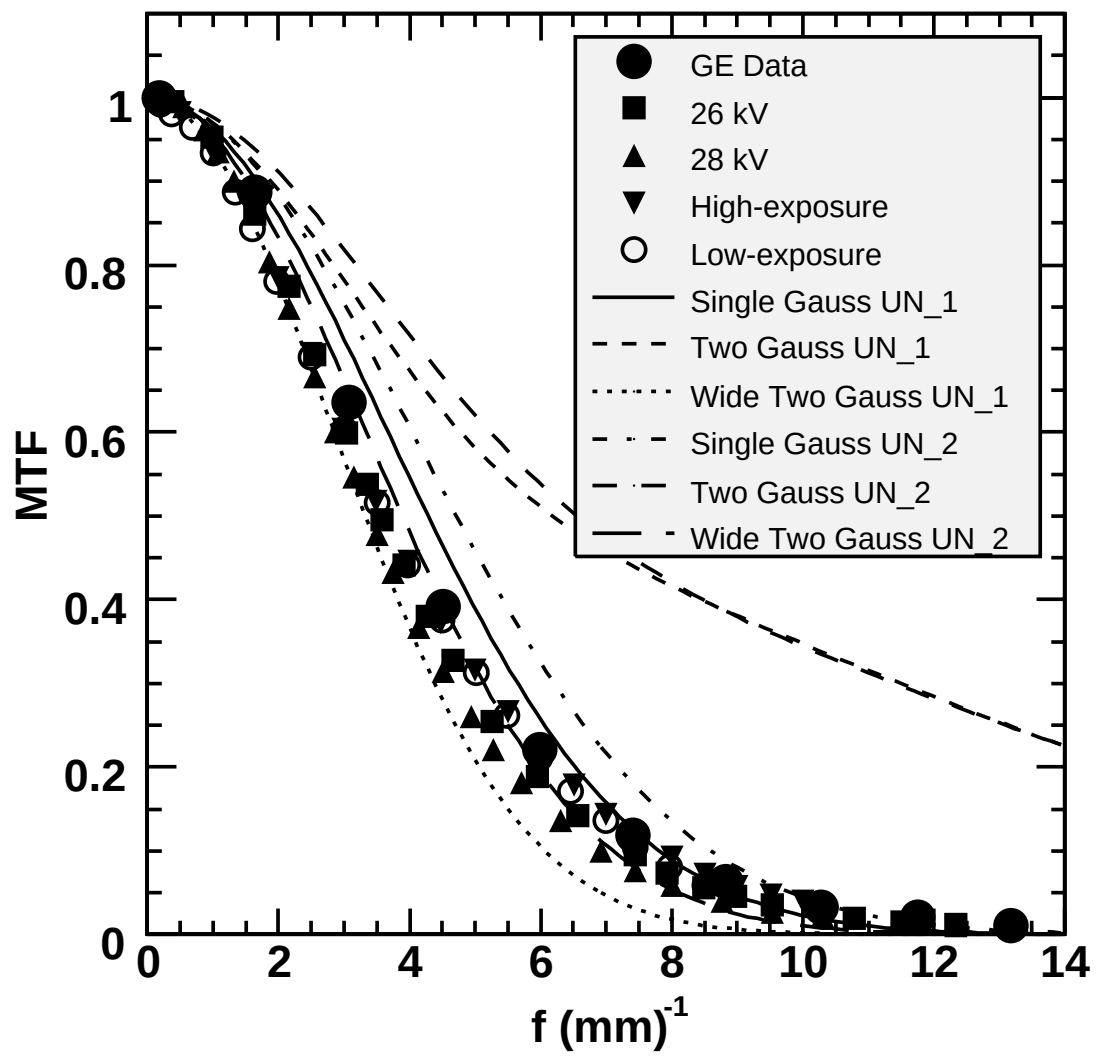

Figure 11. MTF data for GE Senographe 2000D detector obtained in this work, together with the data obtained by traditional methods. Symbols are • GE data obtained at 28 and 30 kV are superimposed [16], ■ and ▲ 26 and 28 kV respectively, ▼ and O high and low exposure respectively [17]. Curves represent the current results for 26 and 28 kV for two detection devices and two different representations of the presampled $PSF^0$. The description of symbols and lines is shown in the inset.